# Water-Vapor Absorption Database using Dual Comb Spectroscopy from 300-1300 K Part II: Air-Broadened $H_2O$, 6600 to 7650 cm$^{-1}$

Scott C. Egbert[1], Keeyoon Sung[2], Sean C. Coburn[1], Brian J. Drouin[2], Gregory B. Rieker[1]

1 – Precision Laser Diagnostics Laboratory, University of Colorado, Boulder

2 – Jet Propulsion Laboratory, California Institute of Technology

## Abstract

We present broadband dual frequency comb laser absorption measurements of 2% $H_2O$ (natural isotopic abundance of 99.7% $H_2^{16}O$) in air from 6600-7650 cm$^{-1}$ (1307-1515 nm) with a spectral point spacing of 0.0068 cm$^{-1}$. Twenty-nine datasets were collected at temperatures between 300 and 1300 K (±0.82% average uncertainty) and pressures ranging from 20 to 600 Torr (±0.25%) with an average residual absorbance noise of 8.0E-4 across the spectrum for all measurements. We fit measurements using a quadratic speed-dependent Voigt profile to determine 7088 absorption parameters for 3366 individual transitions found in HITRAN2020. These measurements build on the line strength, line center, self-broadening, and self-shift parameters determined in the Part I companion of this work. Here we measure air-broadened width (with temperature- and speed-dependence) and air pressure shift (with temperature-dependence) parameters. Various trends are explored for extrapolation to weak transitions that were not covered in this work. Improvements made in this work are predominantly due to the inclusion of air pressure shift temperature dependence values. In aggregate, these updates improved RMS absorbance error by a factor of 4.2 on average, and the remaining residual is predominantly spectral noise. This updated database improves high-temperature spectroscopic knowledge across the 6600-7650 cm$^{-1}$ region of $H_2O$ absorption.

## Introduction

$H_2O$ vapor is a strong absorber of infrared radiation and is found in a host of combustion, atmospheric, and biological systems. Accurate characterization of $H_2O$ vapor absorption, particularly when mixed with air, is crucial for understanding terrestrial radiative processes [1], [2], interpreting spectra from exoplanetary atmospheres [3], [4], and making non-intrusive combustion measurements [5], [6], [7], among a host of other applications. By comparing spectrally resolved absorption measurements against molecular absorption models, it is possible to determine the environmental conditions. For example, because $H_2O$ is an abundant product of hydrocarbon combustion, near-infrared $H_2O$ absorption is commonly used to measure thermodynamics parameters such as temperature [5], [6], [7], [8], [9], [10], [11], [12], concentration [5], [6], [7], [8], [9], [10], [11], [12], pressure [5], velocity [5], [11], [12], and even mass flux [5] in high-temperature reacting systems.

For molecules with discrete absorption transitions, such as $H_2O$, generating accurate, tabulated line-by-line absorption models hinges on two crucial elements: (1) the parametrization of the line shape model to account for the relevant molecular physics under the specific measurement conditions, which includes factors like collisional narrowing, foreign broadening, and temperature dependence, and (2) the tabulation of parameters within a comprehensive database, enabling model calculations across a spectrum of conditions. The failure to adequately address either of these aspects results in an erroneous spectral model and inaccurate calculations.

As described in Part I of this work [13], leading spectral parameter databases, such as HITRAN [14], HITEMP [15], ExoMol [16], and GEISA [17] include tens to hundreds of thousands of well-curated individual absorption transitions. Currently, these databases all employ a Voigt profile (VP) to represent the line shape of air-broadened $H_2O$ absorption in the spectral region of interest. With improving measurement techniques and instrumentation, it has become increasingly evident that the VP model inadequately captures the true absorption line shape of $H_2O$, especially at elevated temperatures [18], [19], [20], [21], [22]. An alternative approach, the Partially-Correlated

Quadratic-Speed-Dependent Hard-Collision (pCqSDHC) model, demonstrates close agreement with high-resolution measurements [20]. At a fraction of the computational cost of the pCqSDHC model, augmenting the VP with quadratic speed dependence (SD) to create the SDVP model has been shown to match the signal-to-noise levels of many measurements [18], [19], [21].

In addition to identifying an appropriate line shape model, many parameters change as a function of temperature, including collisional narrowing and transition center frequency. While research into the temperature dependence of spectral shifts is ongoing, none of the databases described include parameters describing the temperature dependence (TD) of the shift in center frequency. Including this TD parameter was recently shown to reduce Doppler-shift-based velocity errors from 18% using the HITRAN database to 2% using the database presented in this work.

In Part I of this study [13], we updated line centers, line strengths, self-width, and self-shift parameters (with TD and SD of the width) contained in the HITRAN2020 database using 29 measurements spanning the same measurement region. Using these measurements, we updated 17030 absorption parameters for 5986 transitions and identified 574 features not presently catalogued in the database. Building on this updated $H_2O$ database, here we use an additional 29 dual comb spectroscopy (DCS) measurements of absorption of 2% $H_2O$ in air to update the foreign broadening and shift parameters of that same database. These measurements also span 6600 to 7650 cm$^{-1}$ with temperatures ranging from 300 to 1300 K and pressures from 20 to 600 Torr. In this study we measure 7088 absorption parameters for 3366 transitions, updating parameters for air-broadened and shifted $H_2O$ (similarly with TD and SD of the width).

Overall, we see an improvement in RMS residual noise from 0.0025 to 0.0006, averaging across all measurements, with the improvement originating from updates to a balance of both the broadening and shift parameters. The remaining residual is largely due to spectral noise with a small residual due to the TD of SD parameters for the largest features. After computing values for the SDVP database, we removed SD and re-optimized the parameters for use when the Signal-to-Noise Ratio (SNR) or computational bandwidth only requires the VP.

This work is the only broadband (hundreds of cm$^{-1}$), high-temperature (over 1000 K), experimentally derived air-broadened $H_2O$ database known to the authors. Together with the foundation derived from the pure $H_2O$ measurements in Part I, this work presents an empirically-validated $H_2O$ database optimized for high-temperature combustion and exoplanetary measurements in the presence of air.

## Experimental Setup

### Spectrometer

As described in Part I of this work [13], Dual Comb Spectroscopy (DCS) is a promising technique well-suited for high-temperature database applications. It combines several advantageous characteristics of traditional Fourier Transform Spectroscopy (FTS) and Tunable Diode Laser Absorption Spectroscopy (TDLAS). Much like FTS, DCS can capture wide spectra spanning over 1000 cm$^{-1}$ while maintaining high resolution for individual absorption transitions at terrestrial conditions (less than 0.1 cm$^{-1}$). Like TDLAS, DCS utilizes an actively modulated laser light source, effectively isolating the absorption measurement from background radiation due to thermal emission – a concern that can compromise the performance of FTS in high-temperature environments.

Our DCS system [13], [23], [24] consists of two mode-locked frequency comb lasers operating at slightly different repetition rates ($f_r$ = 204 MHz, $\Delta f_r$ = 436 Hz), fully phase locked for continuous coherent averaging [25], [26], [27]. Measurements for Parts I and II of this work were taken concurrently using the same optical setup and spectrometer. We lock one comb tooth from each laser to a narrow linewidth Continuous Wave (CW) reference laser (1.5 kHz linewidth RIO Planex) at 1565 nm, resulting in comb tooth linewidths on the order of tens of kHz [28]. Comparable to [13], the precision of the frequency axis was primarily influenced by the stability of the CW laser, exhibiting a sinusoidal fluctuation with an amplitude of ±5 Hz (50 ppb out of 204 MHz), thus introducing an estimated frequency axis uncertainty of 1.74E-4 cm$^{-1}$ [23]. We actively reference all frequency measurements to a

GPS-disciplined quartz oscillator (Jackson Labs Fury) with 10 ppt stability, contributing negligible frequency drift in comparison to other sources of uncertainty.

The laser pulses generated by each comb are individually amplified and spectrally broadened to include hundreds of thousands of comb modes (wavelengths), spanning the range of 1.3-1.8 μm, evenly spaced at the laser repetition rate (204 MHz or 0.0067 cm$^{-1}$). We then spectrally filter this light to 6600-7650 cm$^{-1}$ using an optical lens and diffraction grating. A slight detuning of the frequency comb repetition rates ($\Delta f_r$ = 436 Hz), gives a unique frequency offset to each successive pair of comb modes. We use a detector to measure the heterodyne beats (interferogram) between the lasers where the RF frequencies present in the interferogram (sampled from 0-100 MHz) correspond to the comb mode pairs of the dual-comb lasers. The amplitude of each measured RF frequency indicates the transmission (or attenuation) of the laser at the corresponding infrared frequency. This process transforms the THz infrared frequencies of successive pairs of comb modes into MHz frequency beats, directly proportional to the magnitude of these comb mode pairs [27], [29]. Like Cole [23], we were careful to avoid measurement distortions due to phase noise [30], detector linearity [31], and cross-phase modulation [32].

Recent advancements in DCS measurements [33] demonstrated a sub-0.1% precision in atmospheric $O_2$ concentration measurements by including a voltage dither with the interferogram signal before digitization. This dither randomized interferogram data points across numerous digitization registers, reducing any potential bias stemming from excessive reliance on registers near the average voltage. Employing the same technique in this study, we applied a 1 MHz, 100 mV$_{pp}$ dither to the detector output before digitization. Notably, no digital filtering was necessary to eliminate the dither signal, as the 1 MHz frequency corresponds to the heterodyne beat note that emerged between comb teeth at 6450 cm$^{-1}$, a region outside the measurement range for this work. Interferograms for each measurement condition were meticulously phase-corrected and coherently averaged over approximately one hour, yielding an average absorbance noise level of 8.0E-4 across the entire spectrum for all conditions under investigation, comparable to the 7.4E-4 absorbance noise from Part I [13].

**Furnace and Spectroscopy Cell**

In Figure 1, we present a schematic of the high-temperature optical cell used in this study. This is the same setup we presented in Part I [13], with the addition of a mixing tank to hold a consistent, pre-mixed charge of air-water gas. The optical setup consists of a 45.7 cm quartz static-sample cell inside of a purged quartz tube at the center of a 106 cm long furnace (CM Furnace) with three independent temperature zones. We double-passed the DCS laser beam along the upper region of the spectroscopy cell, resulting in a cumulative absorbing pathlength of 91.4 ± 0.1 cm. This extended optical pathlength enables us to detect and analyze a multitude of weakly absorbing features, especially at the lower number densities observed at elevated temperatures.

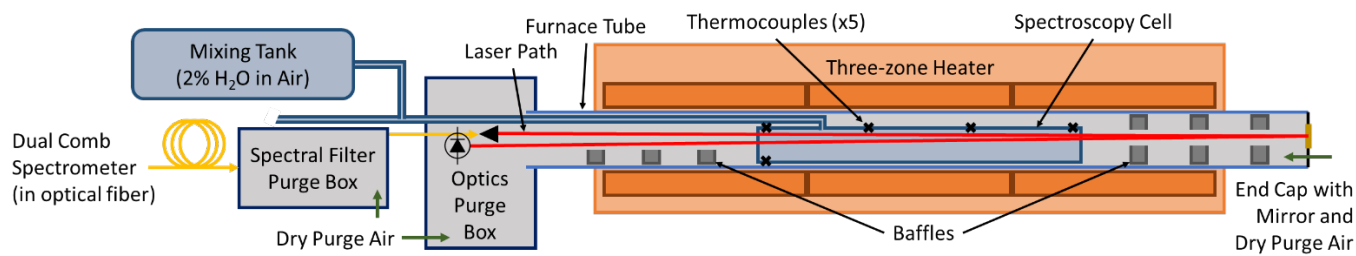

Figure 1: Schematic of the high-temperature furnace used for data collection in [13] and this work. A 45.7 cm spectroscopy cell is centered in a 106 cm three-zone furnace. Baffles were used to reduce convective currents and improve axial temperature uniformity. The 3 cm diameter absorption cell was placed in a 9 cm diameter, 165 cm long quartz tube that was capped at one end and enclosed at the other to allow all free space optics to be purged of background $H_2O$ absorption. The DCS laser beam was passed through a purged spectral filter before being passed twice through the furnace for an effective pathlength of 91.4 cm. A 12 L mixing tank was used to give a consistent mixture of air and $H_2O$ for each measurement.

To mitigate temperature variations arising from convection, baffles were strategically placed at the furnace's periphery. Whenever experimental conditions were altered, the furnace temperature was allowed to stabilize for a minimum of 3 hours. Additionally, a continuous flow of dry air was employed to reduce and stabilize the moisture content from ambient air in the furnace and the grating-based spectral filter.

We measured temperature of the absorption cell in five locations using k-type thermocouples (Omega): four axial positions along the upper section of the spectroscopy cell, positioned closest to the laser beam's path, and an additional radial measurement at the cell's lower section to assess radial temperature variations. Our reported temperature uncertainty includes instrument, temporal, axial, and radial variations, added in quadrature. On average, the temperature uncertainty was 0.82% relative to the measurement value, with a maximum uncertainty of 1.54% at 1300 K. The dominant source of uncertainty at lower temperatures is the thermocouple measurement uncertainty itself, and at higher temperature, axial non-uniformity. Temporal and radial variations remained negligible across all experimental conditions.

Comparable to the companion study presented in Part I [13], high purity, vacuum degassed $H_2O$ (Honeywell CAS 7732-18-5) of natural isotopic abundance (99.7% $H_2^{16}O$) was used for each measurement. For this work, the $H_2O$ was pre-mixed with synthetic air (Airgas Ultra Zero, less than 2ppm of $H_2O$ and 1ppm of $CO_2$+CO) in a stainless-steel mixing tank (12 L) before being measured in the furnace. When preparing a sample, we filled this tank to 17 Torr of pure $H_2O$ ($P_{sat,295K}$ = 19 Torr) before pressurizing to 850 Torr using the synthetic air, targeting a concentration of 2% $H_2O$ in air. We then agitated the stainless-steel ball bearings that had been placed in the tank for 15 minutes to thoroughly mix the air and water before leaving the mixture to fully equilibrate for at least 12 hours before use. Two batches of air-water mixture were used to complete all measurements: one for all 300 and 500 K measurements and a second for 700, 900, 1100, and 1300 K measurements.

Before each measurement, we first flushed the absorption cell down to tens of mTorr, filled it to near the measurement pressure using a fresh charge of air-water mixture, allowed this mixture to sit for nominally five minutes, emptied the cell again, and then filled the cell to the desired pressure for the measurement condition. We used this process to equilibrate the $H_2O$ adsorption in the stainless-steel tubing, stabilizing the water concentration for each measurement. DCS measurements were taken after the total pressure stabilized, indicating the $H_2O$ was in equilibrium with the stainless-steel, a process that took nominally five minutes after filling the absorption cell.

Pressure in the spectroscopy cell was continually monitored using a 1000 Torr Baratron capacitance manometer (MKS-628D) with an uncertainty of 0.25% of the measured value that was calibrated (NCSL Z540.1) immediately prior to data collection. Our reported pressure uncertainty consists of both temporal and instrument uncertainties added in quadrature, averaging 0.25% for all measurements with a max of 0.26% due to minor temporal fluctuations at 1100 K, 40 Torr. The leak rate of the 2-liter absorption cell volume was negligible at less than 0.1 mTorr per hour over a 60-hour test below 10 mTorr. We actively evacuated the spectroscopy cell overnight before each day of data collection to fully remove $H_2O$ vapor from the cell and tubing. A background measurement was then taken before adding $H_2O$ to the cell to characterize the shape of the laser baseline and to quantify background $H_2O$ absorption from the air in the spectral filter and outside of the absorption cell in the furnace.

The test matrix summarizing conditions and respective uncertainties for all measurements is shown in Table 1. In total, 29 spectra were taken, with at least five different pressure measurements at each of five temperatures. Only one measurement (600 Torr) was collected at 1300 K to reduce damage to the quartz from prolonged exposure to $H_2O$ vapor at high temperatures. Additional low-temperature measurements were taken to better characterize strong absorbing features that saturate at lower temperatures and higher pressures (i.e. 8 features at 300 K and 300 Torr have transmission below 15%).

Table 1: Average conditions measured for this work. Temperature uncertainty is dominated by thermocouple instrument uncertainty at low temperatures and a mix of thermocouple and axial uniformity at elevated temperatures. Pressure uncertainty is dominated by pressure gage instrument uncertainty. Additional room temperature measurements were taken at lower pressures to better characterize strongly absorbing features. Only one measurement at 1300 K was taken to avoid quartz/water vapor exposure damage.

| Temperature (K) | Pressure (Torr) |
| --- | --- |
| 295 ± 2 | 20.2±0.1, 40.3±0.1, 59.6±0.1, 80.0±0.2, 119.6±0.3, 159.7±0.4, 319.9±0.8, 599.5±1.5 |
| 505 ± 4 | 40.0±0.1, 79.6±0.2, 162.1±0.4, 320.8±0.8, 601.0±1.5 |
| 704 ± 5 | 40.0±0.1, 80.0±0.2, 160.2±0.4, 319.9±0.8, 599.9±1.5 |
| 901 ± 8 | 40.1±0.1, 80.2±0.2, 159.8±0.4, 319.3±0.8, 599.8±1.5 |
| 1099 ± 11 | 40.1±0.1, 80.2±0.2, 160.0±0.4, 319.8±0.8, 599.5±1.5 |
| 1288 ± 20 | 599.4±1.5 |

We optically calculated $H_2O$ concentration for each measurement using the measured absorption from a combination of 225 isolated features and 26 isolated doublet pairs, all of which were visible and unambiguously isolated from other absorption transitions at all measurement conditions. An SDVP (foreign width with TD, foreign shift with TD, and SD of the foreign width) was fit to each of the features with sufficient SNR, or feature pairs, at every condition using the line strength, center, and self-SDVP values calculated in Part I [13]. We then averaged the concentrations for all 251 features to obtain the $H_2O$ concentration at that measurement condition, as shown in Figure 2. Due to the 60x variation in number density across measurements, not all features were visible at all measurements conditions. As such, only features with less than 10% uncertainty were included in the average, with the 300 K, 20 Torr condition using the fewest (220 absorption features). We estimate $H_2O$ concentration uncertainty as the standard deviation between the concentrations from each of the up to 251 measured features (averaging 2.9% relative to the concentration) and the relative temperature, pressure, and pathlength uncertainty for each measurement, added in quadrature.

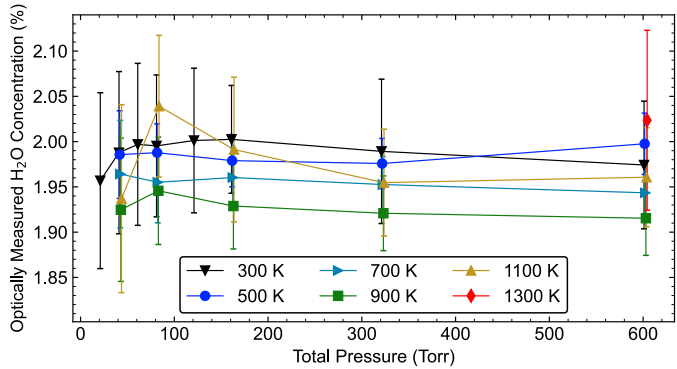

Figure 2: Optically calculated $H_2O$ concentration for each measurement condition (see Table 1), calculated by averaging the concentration obtained from between 220 and 251 isolated absorption features, depending on feature SNR at each condition. Uncertainty sources include fit, temperature, pressure, and pathlength and average 3.1% relative to the concentration.

**Data Preparation**

In this work, we employed the same methodology for processing DCS measurements as presented in Part I [13] and shown in Figure 3 of this work. A series of scans with the optical cell at vacuum conditions were taken each day of testing to both characterize the spectral shape of the laser baseline and to quantify the background water concentration. The raw vacuum scan associated with the 1300 K furnace temperature is shown in the top row of Figure 3 in panels A and B in orange. A two-temperature spectral model was fit to each vacuum scan to quantify high-temperature absorption in the furnace and low-temperature absorption in the spectral filter (see Figure 1). This absorption was then subtracted from the vacuum scan to isolate the laser baseline (teal) before the baseline was low-pass filtered to remove noise and isolate the laser shape (black). As in Part I, the average background $H_2O$ concentration we observed was 292 ppm with variations on the order of 31 ppm between vacuum scans at different conditions.

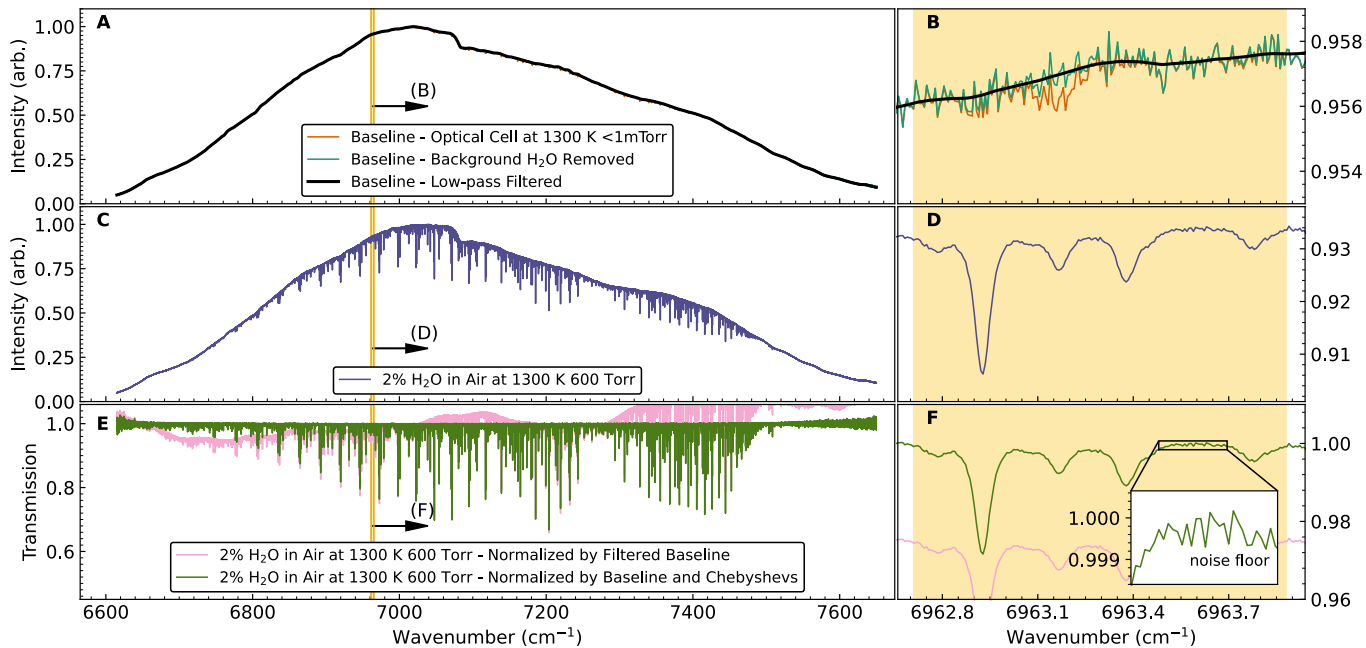

*Figure 3: Measurements and data processing used to prepare transmission data for this work. Panels A and B show the background measurements taken with the sample cell under vacuum to characterize background $H_2O$ absorption (orange), the background scan with molecular absorption quantified and removed (teal), and a smoothed laser transmission baseline (black). Panels C and D show the 1300 K, 600 Torr measurement of 2% $H_2O$ in air (purple). Panels E and F show the same measurement normalized by the smoothed baseline (pink). Additional normalization due to large-scale fluctuations in laser baseline was performed using a 7th order Chebyshev across overlapping 25 $cm^{-1}$ segments to fully normalize the spectrum (green). The inset in panel F highlights the noise floor of the measurement at right (7.5E-4 for the measurement at this condition, slightly below the 8.0E-4 average for all measurements).*

Panels C and D of Figure 3 show the raw measurement as obtained from the DCS at 1300 K and 600 Torr with the comparable laser baseline to the vacuum scan (panel A) clearly visible. We divided the raw measurements by the smoothed vacuum scan to obtain the partially normalized spectrum shown in pink in panels E and F, where the low-frequency oscillations are due to small fluctuations in the laser intensity over time and do not impact individual features. Background $H_2O$ absorption from $H_2O$ outside of the spectroscopy cell, as calculated from the vacuum scan, was then subtracted from the measured absorbance. Returning to transmission, during the fitting routine we used $7^{th}$-order Chebyshev polynomials across 25 $cm^{-1}$ segments of the spectrum to normalize for the remaining low-frequency oscillations, resulting in the green trace presented in Panels E and F. Panel F includes an inset to highlight the measurement noise floor of 7.5E-4 absorbance noise for this specific measurement and 8.0±2.6E-4 for all the measurements in our study. On the left side of this inset, you can observe the distant wings of the feature at 6963.4 $cm^{-1}$.

We excluded data points below the conservative cutoff of 15% transmission from the fitting algorithm due to the high uncertainties associated with high-absorbance/low-transmission data points and the absence of DCS validation under these conditions. For this work, only features at 300 K were affected with 8 features exhibiting high absorbance at 600 Torr, 7 at 320 Torr, 5 at 160 Torr, and 3 at 120 Torr. Because these features absorb light so strongly, they still have excellent SNR in lower pressure measurements. This deliberate choice to omit a few well-documented, low-temperature features at high pressure (rather than limiting the number density of $H_2O$ vapor to avoid high absorbance) enabled us to focus on measuring thousands of high-temperature features, aligning with the primary objective of this research.

## Line Shape Model and Multispectrum Fitting

### Line Shape Model

Line shape models are used to reduce measured spectra into a set of line-by-line absorption model parameters that can be used to predict absorption at a range of conditions. Our dataset includes thousands of individual absorption features with SNR varying by over three orders of magnitude over the noise floor of the measurement. To effectively analyze such a large number of features of different SNR, we applied the multispectrum fitting algorithm initially introduced in [34], a method that has consistently demonstrated its capability to accurately retrieve hundreds of line shape parameters using high-resolution spectra (such as [13], [23], [24], [35], [36], [37]). In multispectrum fitting, all measurements are fit simultaneously to calculate the best values for each parameter. The ability to determine up to five parameters for hundreds of transitions at a time was essential for processing these broadband measurements in a reasonable timeframe and will be described in greater detail in the following section. When reducing spectral measurements into modeled parameters, we used the quadratic speed-dependent (SD) Rautian profile, as defined in Equation 1:

$$F(x, y, a_w, H) = \frac{2}{\pi^{3/2}} \int_{-\infty}^{+\infty} v e^{-v^2} \arctan\left(\frac{x(T,P) + v}{y(T,P)\left(1 + a_w\left(v^2 - \frac{3}{2}\right)\right) + H}\right) dv \qquad (1)$$

Where the dimensionless line shape profile, $F$, is a function of the dimensionless center frequency, $x$, Lorentz width, $y$, and Dicke narrowing coefficient, $H$ [13], [35]. The dimensionless SD parameter, $a_w$, is the ratio of the SD and collisional (Lorentz) broadening. Collisions are normalized by the most probable speed for a given temperature, which was calculated from $v = -4$ to $4$ in 16 steps at each frequency of the transition [35]. Previous analysis has found that SD dominates Dicke narrowing at the temperatures and pressures measured in this work [21], [22], comparable to Part I [13]. Because the matrix-based solver [34] is not designed to simultaneously determine the highly coupled SD and Dicke narrowing, only SD values were analyzed for this study with $H$ set equal to zero, reducing the SD Rautian profile to the SD Voigt Profile (SDVP). Additionally, due to the low concentration of $H_2O$ in this work, air-$H_2O$ SD is the dominant form of SD narrowing. $H_2O$-$H_2O$ SD calculations performed in part I were not included in this study due to limitations of the matrix-based solver.

For an absorbing molecule in air, the dimensionless line center parameter, $x$, from Equation 1 is parameterized as shown in Equation 2:

$$x(T,P) = \frac{\sqrt{\ln(2)}}{\gamma_{Doppler}}\left(\sigma - \sigma_0 - P[X_{self}\delta_{self}(T) + X_{air}\delta_{air}(T)]\right) \qquad (2)$$

Where $\sigma$ is the frequency at which the lineshape is to be determined, $\sigma_0$ is the center frequency of the transition at vacuum, $P$ is the gas pressure, $X_{self}$ is the concentration of the absorbing molecule, $\delta_{self}$ is the collisional self-shift at the gas temperature, $T$, $X_{air}$ and $\delta_{air}$ are the concentration of air and corresponding air pressure shift, and $\gamma_{Doppler}$ is the Doppler width of the transition [35]. Similarly, the parameterization for the dimensionless line width parameter, $y$, is shown below in Equation 3:

$$y(T,P) = \frac{\sqrt{\ln(2)}}{\gamma_{Doppler}} P[X_{self}\gamma_{self}(T) + X_{air}\gamma_{air}(T)] \qquad (3)$$

Where $\gamma_{self}$ is the Lorentz self-width and $\gamma_{air}$ is the air-width at the gas temperature. For both Equations 2 and 3, describing the variation of $\delta_{self}$, $\delta_{air}$, $\gamma_{self}$, and $\gamma_{air}$ with temperature is a non-trivial subject of continued investigation and will be discussed in detail in the following sections.

**Multispectrum Fitting Procedure**

We employed a comprehensive multispectrum fitting approach, utilizing all acquired data collectively to extract optimal fits for each feature under diverse conditions [34]. To accelerate each processing iteration, we partitioned the spectral data into 58 segments averaging 20±2 cm$^{-1}$ wide, excluding regions with sparse H$_2$O absorption near 6600 and 7650 cm$^{-1}$, comparable to part I [13]. We tailored these bin widths to ensure complete isolation of spectral features within each bin. Recognizing the potential slow drift in laser intensity between vacuum and measurement scans, we employed an additional baseline normalization strategy for each bin using 7$^{th}$ order Chebyshev polynomials, adding an extra 2 cm$^{-1}$ buffer on both sides of each bin to mitigate edge-related artifacts. A 7$^{th}$ order Chebyshev polynomial spanning nominally 24 cm$^{-1}$ had negligible impact on spectral features, which exhibited a maximum width on the order of 1.2 cm$^{-1}$ for the largest of overlapping features.

As discussed in Part I of this work for pure H$_2$O, identifying how the spectral widths and shifts shown in Equations 2 and 3 change with temperature is an essential part of an accurate spectroscopic database. Furthermore, when measuring molecules with congested absorption spectra, as is the case for air-broadened H$_2$O in this region, appropriate TD relationships can help discriminate among overlapping transitions by combining information from measurements at multiple temperatures.

The single power law (SPL) relationship, as shown in Equations 4 and 5 for the air-broadened Lorentz width and air pressure shift, is a common TD parameterization that was used in this multispectrum fitting routine, comparable to the measurements presented in Part I of this work for self-widths and shifts.

$$\gamma_{air}(T) = \gamma_{air}^0 \left(\frac{T_{ref}}{T}\right)^{n_{\gamma,air}} \tag{4}$$

$$\delta_{air}(T) = \delta_{air}^0 \left(\frac{T_{ref}}{T}\right)^{n_{\delta,air}} \tag{5}$$

Here, $\gamma_{air}(T)$ is the observed Lorentz width due to collisions between the absorbing molecule and air at the gas temperature, $T$, as shown in Equation 3. $\gamma_{air}^0$ is the air-broadened half-width at the reference temperature, $T_{ref}$, and $n_{\gamma,air}$ is the SPL TD exponent. In Equation 5, $\delta_{air}(T)$ is the collisional air pressure shift, $\delta_{air}^0$ is the air pressure shift at the reference temperature, and $n_{\delta,air}$ is the SPL TD exponent.

A Double Power Law (DPL) model, where two sets of the SPL TD model are added together, was proposed by Gamache and Vispoel [38] and explored by Stolarczyk et al. [39] where it showed good agreement to theoretical calculations of H$_2$O widths and shifts over broad temperature ranges, such as 100-3000 K. Due to limitations inherent to processing thousands of transitions that are often only visible at a subset of the six measured temperatures, the four parameter DPL model has not been integrated into the multispectrum fitting routine used in this work. The impact of using SPL TD for widths and shifts will be explored in greater detail in the next section.

To estimate initial values for $a_w$, we analyzed 127 isolated transitions where statistical fit uncertainties for $a_w$ and $n_{\gamma,air}$ were below ± 0.007 while floating other width and shift parameters and found an average value for of $a_w$ 0.130 ± 0.022. While we will show in the Results and Discussion that $a_w$ is not as constant for H$_2$O-air collisions as was found in Part I for H$_2$O-H$_2$O collisions, 0.13 was used as an initial estimate for $a_w$ due to the lack of $a_w$ parameters in the HITRAN database. In this analysis of the most isolated features, we did not find notable trends in $n_{\delta,air}$ that could be reliably extrapolated to initial values and instead chose to set initial values of all $n_{\delta,air}$ parameters to 1 (proportional to P/T).

Our work is built on the database we developed in Part I using measurements of pure $H_2O$ [13]. In that work, we calculated line strengths, centers, self-widths, and self-shifts using a similar procedure to what we used for the air-$H_2O$ measurements. We use a similar, systematic processing procedure to obtain comparable results across the 1050 cm$^{-1}$ wide spectrum by establishing an upper limit for statistical uncertainties for each parameter. For each spectral feature, we incrementally float additional parameters until the predetermined uncertainty limit is no longer met for all of the evaluated parameters associated with that feature, iterating ten times to ensure stable convergence. The sequential steps used in this work are outlined below:

1. We floated parameters to minimize a large residual that could disrupt calculations for nearby features, typically width parameters of larger features, although occasionally spectral shifts.
2. Width parameters were then floated for all features above the measurement noise floor, beginning with $\gamma_{air}^0$, then $n_{\gamma,air}$, and finally $a_w$. Parameters that could not be determined to within ±0.12, ±0.13, and ±0.10, respectively for $\gamma_{air}^0$, $n_{\gamma,air}$, and $a_w$ were only floated if the updated value stably reduced model residuals and were evaluated on a case-by-case basis.
3. Finally, shift parameters were floated for features meeting all prior uncertainty limits. Uncertainty limits for $\delta_{air}^0$ and $n_{\delta,air}$ were ±0.005 cm$^{-1}$ and ±0.13, respectively.

We repeated this operation for all 58 bins across the 1050 cm$^{-1}$ measurement range. In some instances, features overlapped in a way that prevented us from identifying unique parameters for the underlying transitions at the conditions measured. Work by Toth and Brown have identified families of features based on quantum assignment that share width coefficients [40], [41]. Like Schroeder et al. [36] and Part I of this work [13], we constrained the width coefficients ($\gamma_{air}$, $n_{\gamma,air}$, and $a_{w,air}$) of doublet pairs to be equivalent. We did not constrain shift relationships ($\delta_{air}$ and $n_{\delta,air}$) due to the ambiguous relationship between doublet pairs.

**Temperature Dependence Model**

In this section we will first describe common TD parameterizations as alternatives to the Single Power Law (SPL) and then present measurements of $\gamma_{air}$ and $\delta_{air}$ as a function of temperature for isolated transitions to explore the limitations and benefits of the SPL model used for in the multispectrum fits presented in this work. This parallels analysis performed in Part I for pure $H_2O$ where the SPL TD was shown to fit the data to the noise floor of the measurement.

Due in part to the anharmonicity of $H_2O$, $\gamma_{air}$ and $\delta_{air}$ can exhibit complicated behavior as a function of temperature. For $\delta_{air}$ in particular, this can include changes from positive to negative spectral shifts, which cannot be represented using SPL TD. Consequently, simple, linear TD models have also been explored to represent $\delta_{air}$, as shown in Equation 6:

$$\delta_{air}(T) = \delta_{air}^0 + \delta'(T - T_{ref}) \tag{6}$$

Where $\delta'$ is the TD slope coefficient. While the linear TD model can represent changes in $\delta_{air}$ sign, it is not well suited to observed trends in TD and extrapolates poorly over temperature ranges as low as 100 K [39].

Recent analysis by Gamache and Vispoel [38] exploring higher fidelity parameterizations for $\gamma_{air}$ and $\delta_{air}$ have presented a Double Power Law (DPL) TD relationship, as shown in Equations 7 and 8, to more accurately represent the expected TD at all gas temperatures.

$$\gamma_{air}(T) = \gamma_{air}^a \left(\frac{T_{ref}}{T}\right)^{n_{\gamma,air}^a} + \gamma_{air}^b \left(\frac{T_{ref}}{T}\right)^{n_{\gamma,air}^b} \tag{7}$$

$$\delta_{air}(T) = \delta_{air}^a \left(\frac{T_{ref}}{T}\right)^{n_{\delta,air}^a} + \delta_{air}^b \left(\frac{T_{ref}}{T}\right)^{n_{\delta,air}^b} \tag{8}$$

Where the combination of $\gamma_{air}^{a}$ and $\gamma_{air}^{b}$, each with corresponding temperature dependencies $n_{\gamma,air}^{a}$ and $n_{\gamma,air}^{b}$, in Equation 7 are used to represent $\gamma_{air}$ with more flexibility than SPL TD. Equation 8 allows for both positive and negative values of $\delta_{air}$ as a function of temperature.

The DPL is an extended model to capture the temperature dependence trends across broad temperature ranges, e.g. 0 - 3000 K, as shown for the theoretical calculations in [38], [39]. As was shown in Part I of this work, fitting DPL TD to finite measurements over a limited temperature range is challenging. With only six measured temperatures between 300-1300 K, it can be unclear if the four DPL parameters determined from these measurements are overfitting the measurement. Additionally, the vast majority of the 3366 observed transitions in this work are only present above the measurement noise floor in a subset of these six measured temperatures.

To better understand TD behavior for the measurements presented in this work, 221 out of the 3366 observed transitions were identified that are (1) present above the noise floor at all six measured temperatures and (2) have no neighboring transitions that contribute (interfere) meaningfully with the measured absorption at any of the measured temperatures. For these 221 transitions, $\gamma_{air}$, $\delta_{air}$, and $a_w$ were retrieved at each temperature using all measurements at that temperature (typically five pressures, with eight pressures at 300 K and one at 1300 K, see Table 1). This process was confirmed to fit the data to the noise floor of the measurement. Models calculated using the lineshape parameters from the multispectrum fitting routine described in the previous section as well as HITRAN data, where available, are also compared to these temperature specific retrievals.

Temperature specific parameters (left) and transmission data (right) for the 2ν$_2$+ν$_3$ (4$_{2,2}$ ← 3$_{2,1}$) transition are shown in Figure 4, where rotational quanta are listed using J$_{Ka,Kc}$ notation. Retrieved values for $\gamma_{air}$ (panel A), $\delta_{air}$ (B), and a$_w$ (C) are shown on the left, each as a function of temperature. The HITRAN prediction for $\gamma_{air}$ (using SPL TD) is shown in panel A in red with the uncertainty shaded. A multispectrum fit of all measurements in this work (also using SPL TD) is shown in green, with a shaded uncertainty. A DPL TD model was fit to $\gamma_{air}(T)$, weighted by uncertainty, and is shown in blue. Similar results are shown in panels B and C for $\delta_{air}$ and a$_w$, respectively, noting that HITRAN does not include parameters for $n_{\delta,air}$ or SD. Measured transmission data for the highest pressure measured (600 Torr) is shown at the top right in panel D, with the residual of the measurement minus a model computed using the HITRAN database and the updated database from the MultiSpectrum Fit in This Work (MSF TW) on the lower right in panels E and F, respectively.

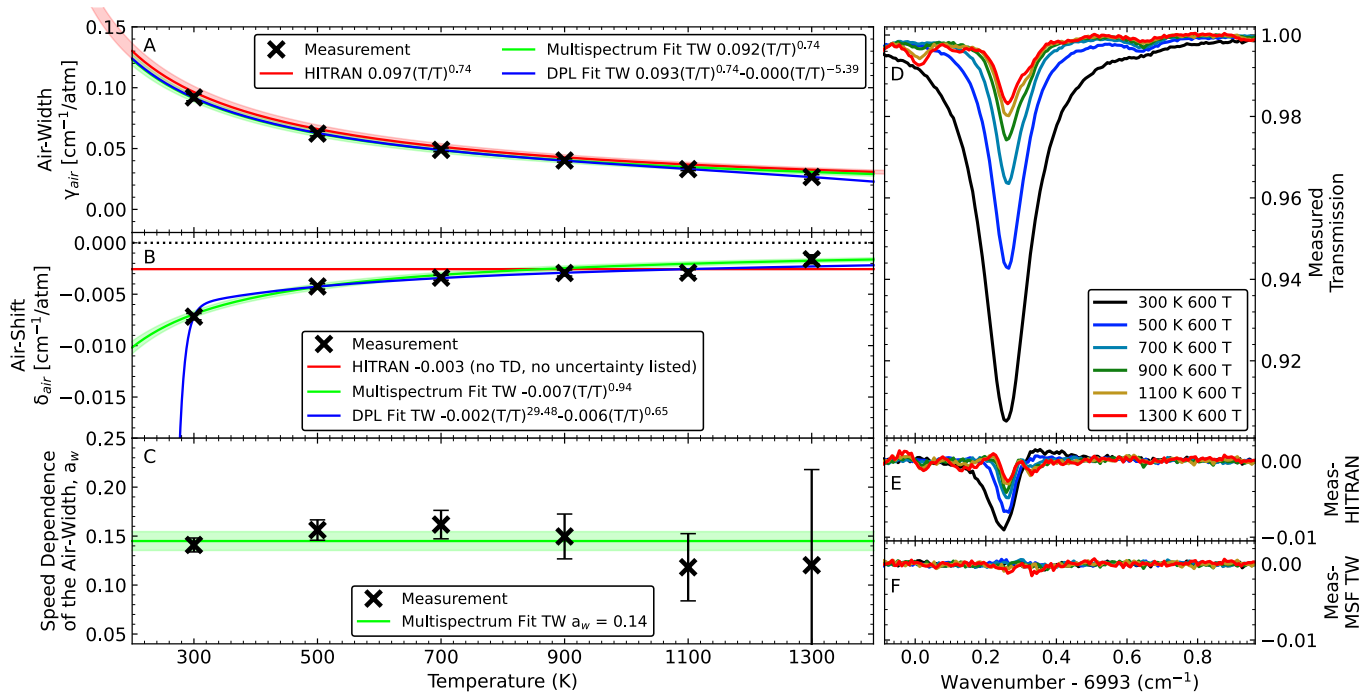

*Figure 4: Left: $\gamma_{air}$ (panel A), $\delta_{air}$ (B), and $a_w$ (C) for the $2\nu_2 + \nu_3$ ($4_{2,2} \leftarrow 3_{2,1}$) transition as a function of temperature. Measured values at each temperature are shown in black, with database values from HITRAN2020 (red, no TD for $\delta_{air}$), measured values using a multispectrum fit (green), and a DPL fit using the measured value at each temperature (blue). Right: six of the 29 transmission measurements that were used to calculate the measurement values (D), the residual when comparing against a model using HITRAN parameters (E) and the MultiSpectrum Fit of This Work (MSF TW) with SPL TD parameters (F), respectively. Note the weak absorption at 1300 K increasing the uncertainty of the retrieved $a_w$ value.*

For both $\gamma_{air}$ and $\delta_{air}$, the SPL parameterization reasonably represents the TD of this transition for the temperature range measured in this work, a trend observed in almost all (over 96%) of the isolated transitions studied. As highlighted in Figure 4, panel A, HITRAN values for $\gamma_{air}^0$ and $n_{\gamma,air}$ generally match measurements of other isolated transitions in this analysis. Similarly, $a_w$ (panel C) for this transition is nominally constant, as was assumed using the parameterization shown in Equation 1.

For the transitions where $\delta_{air}^0$ and $n_{\delta,air}$ were calculated using the multispectrum fitting criteria described previously (77% of the 221 transitions), only a small fraction (~4%) of observed $\delta_{air}(T)$ trends were better represented with a fit using DPL TD across the measured temperature range. For these transitions where SPL TD was measured, the improvements compared to DPL TD in this temperature range were generally small. This does not imply that SPL and DPL parameterizations both adequately represent $\delta_{air}$ TD at all temperatures, only that the upper limit for multispectrum fit uncertainties (±0.005 cm$^{-1}$ and ±0.13, for $\delta_{air}^0$ and $n_{\delta,air}$ respectively) successfully filtered out most transitions where SPL TD gave poor results. This contrasts with Part I for pure H$_2$O, where the low pressures (max pressure of 16 vs 600 Torr) complicated efforts to identify and reject transitions where SPL TD resulted in poor fits. Furthermore, the DPL parameterization appears to overfit the measured $\delta_{air}(T)$ datapoints in Figure 4, panel B, likely giving unreasonable extrapolations below 300 K, where the absorption would be the strongest for this transition.

Figure 5 shows $\gamma_{air}$, $\delta_{air}$, and $a_w$ calculated as a function of temperature for the $\nu_1 + \nu_3$ ($12_{1,11} \leftarrow 13_{1,12}$) transition, with the transmission data again shown on the right. For this high-J transition, both the HITRAN and multispectrum fit values for $\gamma_{air}^0$ at 300 K do not match the measured value shown in panel A. We note that for this type of high-J transition, the discrepancy has no discernable effect on the 300 K residual shown in panel F due to the weak absorption at lower temperature. The DPL parameterization provides the best fit for $\gamma_{air}$ at 300 K. Both DPL and SPL TD provide reasonable predictions above 500 K where the strongest absorption is observed. A similar trend was observed in approximately 3% of the 221 isolated transitions analyzed.

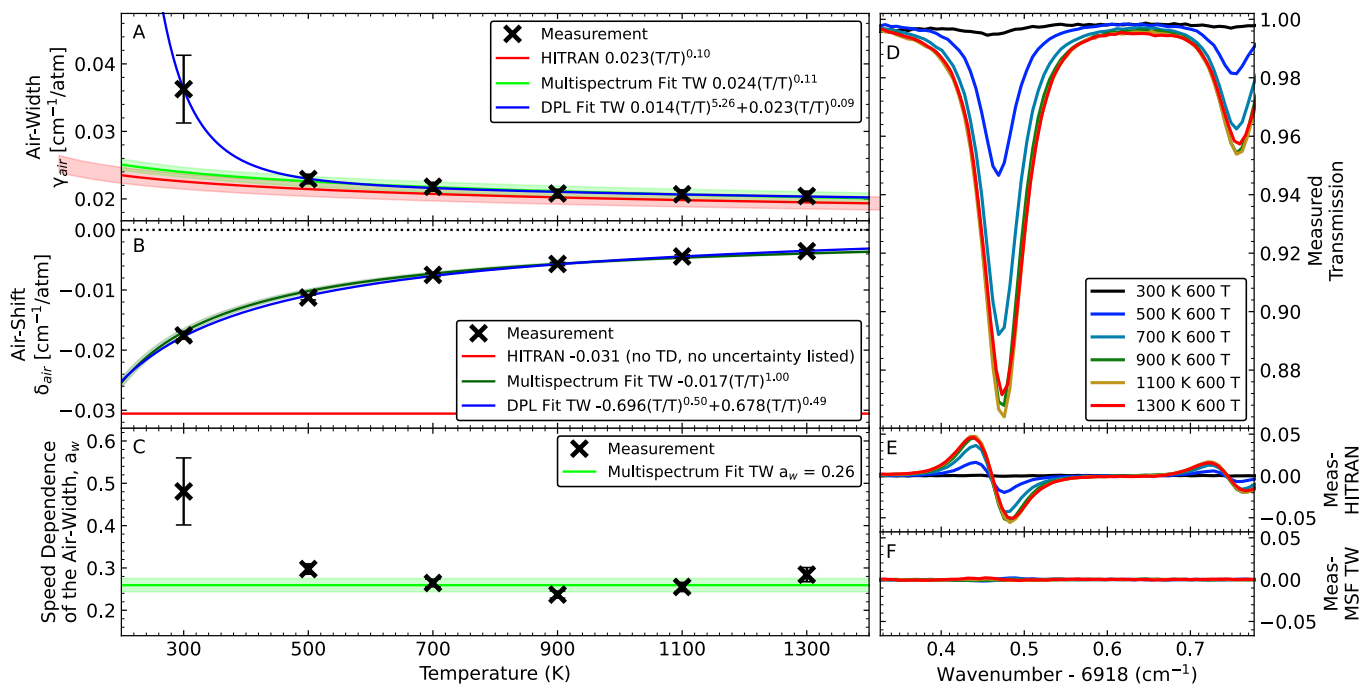

*Figure 5: Left: $\gamma_{self}$ (panel A), $\delta_{self}$ (B), and $a_w$ (C) for the $v_1 + v_3$ ($12_{1,11} \leftarrow 13_{1,12}$) transition as a function of temperature. Measured values at each temperature are shown in black, with database values from HITRAN2020 (red, no TD), measured values using a multispectrum fit (green), and a DPL fit using the measured value at each temperature (blue). Right: six of the 29 transmission measurements that were used to calculate the measurement values (D), the residual when comparing against the HITRAN model € and the MultiSpectrum Fit of This Work (MSF TW) with SPL TD parameters (F), respectively. Note the weak absorption at 300 K increasing the uncertainty of the retrieved $a_w$ value.*

The use of SPL TD for $\delta_{air}$, with an updated value for $\delta^0_{air}$ compared to HITRAN, improves the residual between measurement and model (panel E vs F). While more SD variation is observed for this transition, the majority of the $a_w(T)$ parameters, particularly those at conditions with sufficient SNR to impact the residual (i.e. above 500 K), match the nominal 0.26 measured value when using the multispectrum fit.

The two transitions shown exhibit behavior commonly observed in the other 221 isolated transitions that were strong enough to measure at all six temperatures: (1) when $\gamma_{air}$ TD calculations benefited from fits using the DPL parameterization (~3% of the transitions), SPL TD gave reasonable estimates of parameters at temperatures where the absorption was the strongest and (2) for both $\gamma_{self}$ and $\delta_{self}$, when SPL fit parameters gave similar results to DPL fit parameters within the measured temperature range (~96% of transitions), the DPL parameterization was prone to overfitting, providing questionable extrapolations outside of the measured temperature range. These trends are entirely due to the limited quantity and range of temperatures inherent to physical measurements in a furnace without a cryostat, such as those presented in this work, and are in no way indicative of general DPL suitability for TD parameterization.

With a focus on generating a database optimized for ambient to high-temperature measurements, the authors view the SPL TD parameterization as an effective tool to extract information from absorption features with overlapping transitions, a common occurrence for $H_2O$ in this region, such that a model generated using the database matches observed absorption. From this analysis we concluded that the SPL TD was able to achieve these aims – high fidelity within the measurement range, plausible predictions outside of the measurement range, and maximum improvement for overlapping transitions and transitions only visible at a subset of temperatures. For this reason, and because it would not be possible to fit DPL to most of the 3366 transitions measured, we focused our efforts on SPL parameterizations for both widths and shifts. A companion study will use this data to further explore DPL parameters for select, high-SNR transitions.

In addition to TD of $\gamma_{air}$ and $\delta_{air}$, the assumption that $a_w$ was constant with respect to $\gamma_{air}$ has the potential to impact the fidelity of retrievals. As shown in Figure 4 and Figure 5, $a_w$ is found to be generally constant for the transitions measured at these temperatures when parameterizing $\gamma_{air}$ using the SPL. Among all 221 transitions, $a_w$ varied by nominally ±5.2%, slightly more than the ±3.9% observed for pure H$_2$O absorption.

As described by Lisak et al. in Equation 20 of [42] and Ghysels et al. in Equation 13 of [43], $a_w$ for an absorbing gas in air is postulated to only be a function of $n_{\gamma,air}$, without a unique TD, as described by Equation 9:

$$a_w = \frac{2}{3}(1 - n_{\gamma,air})\frac{\alpha}{1+\alpha} \qquad (9)$$

Where $\alpha$ is the mass ratio of the perturbing vs absorbing molecule. In exploring this relationship for H$_2$O in air, we recognize that the applicability of Equation 9 is inherently coupled to $\gamma_{air}$ SPL TD through $n_{\gamma,air}$ and that air is a gas mixture predominantly composed of two molecules with slightly different mass. The goal of this analysis is to identify additional methods to extrapolate $a_w$ beyond the transitions directly measured in this work. Similar to the studies described by Hartmann et al. [44], however, we did not observe this relationship for pure H$_2$O.

## Results and Discussion

The transmission spectrum shown previously in Figure 3 is presented again in the top row of Figure 6 with the addition of residuals between the measurement and the HITRAN2020 database (HT, second row), the updated database in This Work (TW) using the VP (third row), and the database using the full SDVP (bottom row). As shown in Panel A, the SDVP database best matches the 1300 K measurement, followed by the updated VP database. Root Mean Square (RMS) residual error decreases from 0.0047 (HT) to 0.0010 (VP) and 0.007 (SDVP) between 6750-7501 cm$^{-1}$ for the condition shown (with average RMS for all measurements improving from 0.0025 to 0.0008 and 0.0006 for HT, VP, and SDVP, respectively.

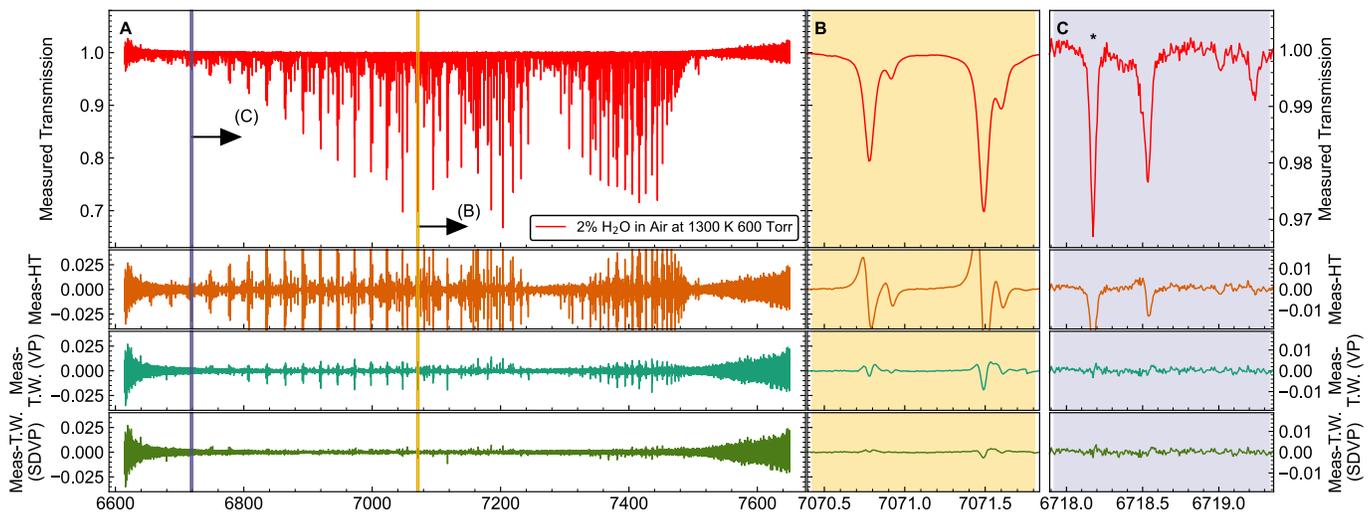

*Figure 6: Normalized, measured transmission spectrum of 2% H$_2$O in air at 1300 K and 600 Torr over a 91.4 cm pathlength (top) with the original residual with respect to HITRAN2020 (HT, second row), the residual from the multispectrum fit employed in This Work (TW) and the VP (third row), and the SDVP (bottom). The entire measured bandwidth is shown in Panel A, with a representative region of strong absorption features shown in Panel B using the same y axis. Panel B highlights the improvement in spectral shifts in the updated database along with the importance of SD between the VP and SDVP. Weakly absorbing features, including a feature that was not included in the HITRAN2020 database (6718.2 cm$^{-1}$, marked with \*), are shown in Panel C. Due to the limited SNR for these weak features, minimal difference is seen between the VP and SDVP parameterizations.*

Zooming in on strong absorption features, in Panel B we see that the largest improvements to the database were spectral shifts, in particular at elevated temperatures. TD shift values are not presently available in HITRAN and, as described by Gamache and Vispoel [38] and Stolarczyk et al. [39], would likely be poorly described by the linear pressure shift TD model currently used in the HITRAN database format at these temperatures. The improvement of feature width between the VP and SDVP is highlighted in the bottom two panels. A small residual is still visible at

7071.5 cm$^{-1}$. After separating the measurements by temperature and evaluating shifts and widths for all measurements at a given temperature, we determined that this lingering residual is due to the difference in TD between the Lorentzian and SD width that is not captured when using a$_w$ for SD.

We highlight the improvements by showing a sample of weakly absorbing transitions in Panel C of Figure 6. As described in Part I, the feature depicted at the left (6718.2 cm$^{-1}$, marked with *) is not presently included in the HITRAN database. Due to the reduced SNR of these weakly absorbing features, SD has minimal impact on the observable residual.

We summarize the quantity of updated parameters from this work, grouped by vibrational band catalogued in the HITRAN2020 database in Table 2. The complete line-by-line database is included in the supplementary material of this paper. In total, we updated at least some parameters for 3347 unique transitions, 99.4% of which originate from main isotopologue H$_2$$^{16}$O (99.7% at natural abundance). Throughout the subsequent sections, we share more detailed analysis and discussion for each of the parameters described in Table 2.

Table 2: Summary of the number of features analyzed for various vibrational bands and isotopologues of H$_2$O visible in this study. Uncatalogued features that were observed were assumed to be of the main isotopologue (H$_2$$^{16}$O). An iterative fitting routine was used in the column order shown, first fitting line center and strengths, followed by width parameters, and finally shift. When fit uncertainties surpassed certainty limits, that value was not updated. An additional 62 features not included in these totals were determined to have line strengths lower than those listed in HITRAN2020. 40 uncatalogued features were also identified with line strengths too weak to be reliably retrieved.

| Vibrational Band | Max J" | Air-Width | Temperature-Dependence of the Air-Width | Speed-Dependence of the Air-Width | Air-Shift | Temperature-Dependence of the Air-Shift |
|---|---|---|---|---|---|---|
| **H$_2$$^{16}$O** | | | | | | |
| 101 ← 000 | 22 | 880 | 524 | 446 | 404 | 254 |
| 200 ← 000 | 20 | 627 | 336 | 249 | 223 | 140 |
| 021 ← 000 | 20 | 463 | 271 | 214 | 157 | 87 |
| 111 ← 010 | 19 | 394 | 88 | 62 | 59 | 22 |
| 031 ← 010 | 18 | 218 | 39 | 21 | 1 | 0 |
| 002 ← 000 | 20 | 184 | 31 | 18 | 9 | 6 |
| 120 ← 000 | 19 | 125 | 34 | 10 | 3 | 0 |
| 201 ← 100 | 12 | 93 | 0 | 0 | 0 | 0 |
| 210 ← 010 | 17 | 90 | 3 | 2 | 0 | 0 |
| 121 ← 020 | 14 | 66 | 0 | 0 | 0 | 0 |
| 041 ← 020 | 13 | 47 | 0 | 0 | 0 | 0 |
| 300 ← 100 | 11 | 47 | 2 | 0 | 0 | 0 |
| 102 ← 001 | 11 | 32 | 0 | 0 | 0 | 0 |
| Other Bands | 20 | 37 | 3 | 2 | 2 | 0 |
| Uncatalogued | | 44 | 0 | 0 | 0 | 0 |
| Subtotal | | 3347 | 1331 | 1024 | 858 | 509 |
| **H$_2$$^{18}$O** | | | | | | |
| 101 ← 000 | 17 | 19 | 0 | 0 | 0 | 0 |
| Total | | 3366 | 1331 | 1024 | 858 | 509 |

**Air-Broadened Half-Width**

Fit results for the 3322 air-broadened half-widths, $\gamma_{air}^0$, measured in this work are shown in panel A of Figure 7 as a function of J" with coloration based on the lower state energy, E". A slight offset of 0.9 Kc"/J" is added to the x-axis to highlight the change of $\gamma_{air}^0$ through the oblate (Kc" = 0) to prolate (Kc" = J") progression. Transitions not catalogued in HITRAN that were identified in Part I of this work are not included on this plot due to the lack of quantum assignments but are included in the supplemental material. HITRAN values for the corresponding transitions are shown in panel B, with the change between this work and HITRAN in panel C. In panels A and C we include measurement uncertainties from measurements in this work, as discussed at the end of this section. HITRAN uncertainties are shown in panel B, with arrows used to indicate unbounded values corresponding to transitions assigned uncertainty codes 3 (over 20%) and 2 (average or constant value). Insets in panels A and B highlight the variation in $\gamma_{air}^0$ as a function of Kc" for J" = 9.

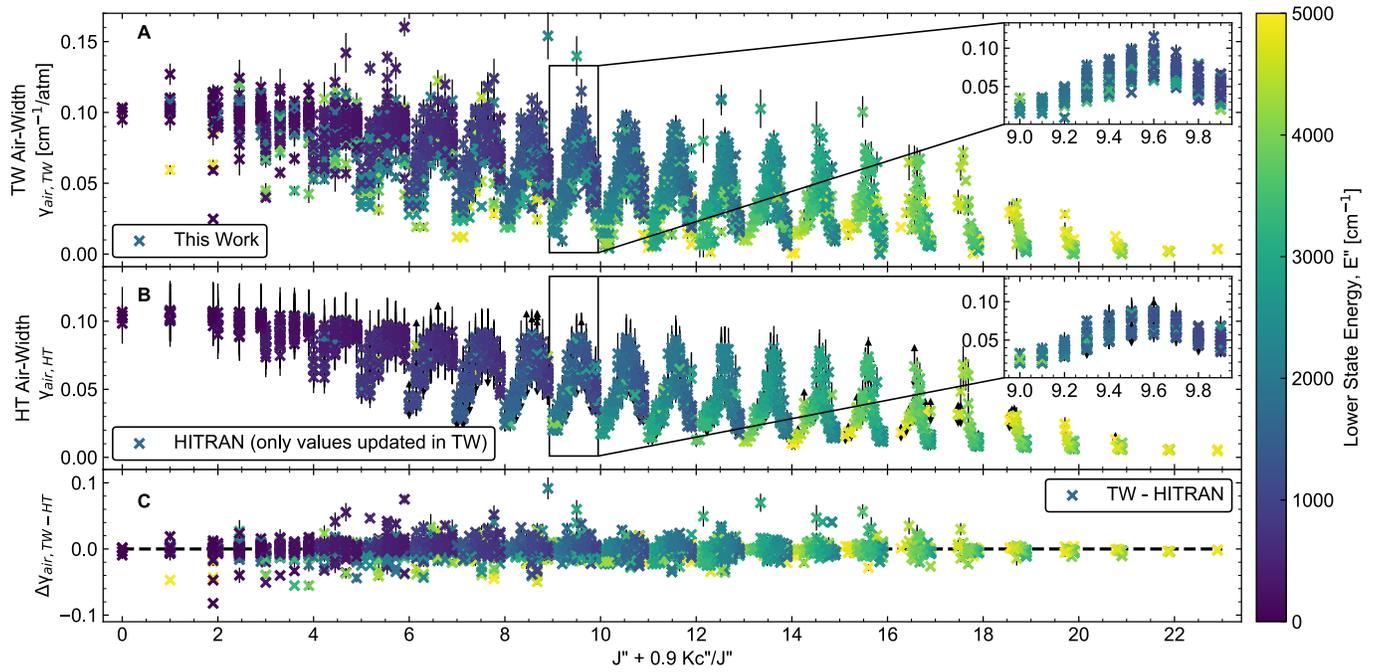

*Figure 7: Measured (panel A) and HITRAN (panel B) air-broadened width values, $\gamma^0_{air}$, as a function of rotational quantum number J" with a small Kc"/J" offset. Insets in the upper right highlight the variation of $\gamma^0_{air}$ as a function of Kc" at J" = 9. Panel C shows the difference in $\gamma_{air}$ between this work and HITRAN.*

In aggregate, we see scattered, though unbiased, agreement between this work and HITRAN, with a linear correlation between the two of $\gamma^0_{air,TW} = -0.0016 + 1.01\gamma^0_{air,HT}$ (R² = 0.88, not shown in Figure 7). The HITRAN values for the majority of updated features (69%) originate from approximations of semi-empirical coefficients from section 4.2.2 in Jacquemart et al. [14], [45] with an additional 22% of HITRAN values derived by Gamache using the Complex Robert Bonamy formulation [46] (see $\gamma^0_{air}$ ref. 60 in [14]). We observe the best agreement between this work and the measurements of Toth [47], [48] (8% of the updated transitions), with -6±7% difference when comparing against $\gamma^0_{air,SDVP}$. The offset with respect to Toth's measuremensts is reduced when comparing against our retrievals using only the Voigt profile, $\gamma^0_{air,VP}$, with a mean difference of -2±8%. On average, our retrieval for $\gamma^0_{air,VP}$ (not shown) was 3% smaller than $\gamma^0_{air,SDVP}$ for the same transitions. Note that the HITRAN database does not include SD effects on spectral widths. These relationships highlight the importance of both semi-empirical calculations to compute large quantities of spectral parameters with general accuracy and reference measurements of individual transitions for high-fidelity retrievals.

Total uncertainty for each transition in this work was estimated using the method of Cole et al. [23], which is to sum the uncertainties in quadrature, including statistical (averaging ±0.003 or ±4.5%), pressure (±0.26%), concentration (±2.9%), and temperature (per Ref. [49]) uncertainties.

**Temperature-Dependence (TD) of the Air-Width**

In Figure 8 we show the single power law TD of the 1331 air-width values we measured in this work, $n_{\gamma,air}$, using the same adjusted J" x-axis as Figure 7. Comparable to Figure 7, panels A, B, and C show $n_{\gamma,air,TW}$ measured in this work, $n_{\gamma,air,HT}$ in the HITRAN database, and the difference between $n_{\gamma,air,TW}$ and $n_{\gamma,air,HT}$, respectively. Over 98% of the HITRAN features updated in this work are assigned HITRAN uncertainty code 2, indicating that $n_{\gamma,air,HT}$ is an average or estimate, as opposed to being assigned a bounded uncertainty (i.e. less than 10%). In spite of this, we observe similar structure between this work and HITRAN, as shown in the insets for both values of $n_{\gamma,air}$. The majority of transitions with J">15 are strongly absorbing doublets where $n_{\gamma,air}$ could be influenced by small errors in center frequency spacing between the line pair or by line mixing. Line-mixing effects are not explored in this study due to the congested nature of H$_2$O absoprtion in this region and the large number of transitions measured. We note that seven high-J" outlier values of $\Delta n_{\gamma,air,TW-HT}$ in panel C were cropped from the image to better highlight the change of the other 1326 transitions (J" = 12, 12, 15, 17, 17, 18, 18). These transitions are included in panel A and the supplemental materials. All but one are indistinguishably overlapping doublet pairs.

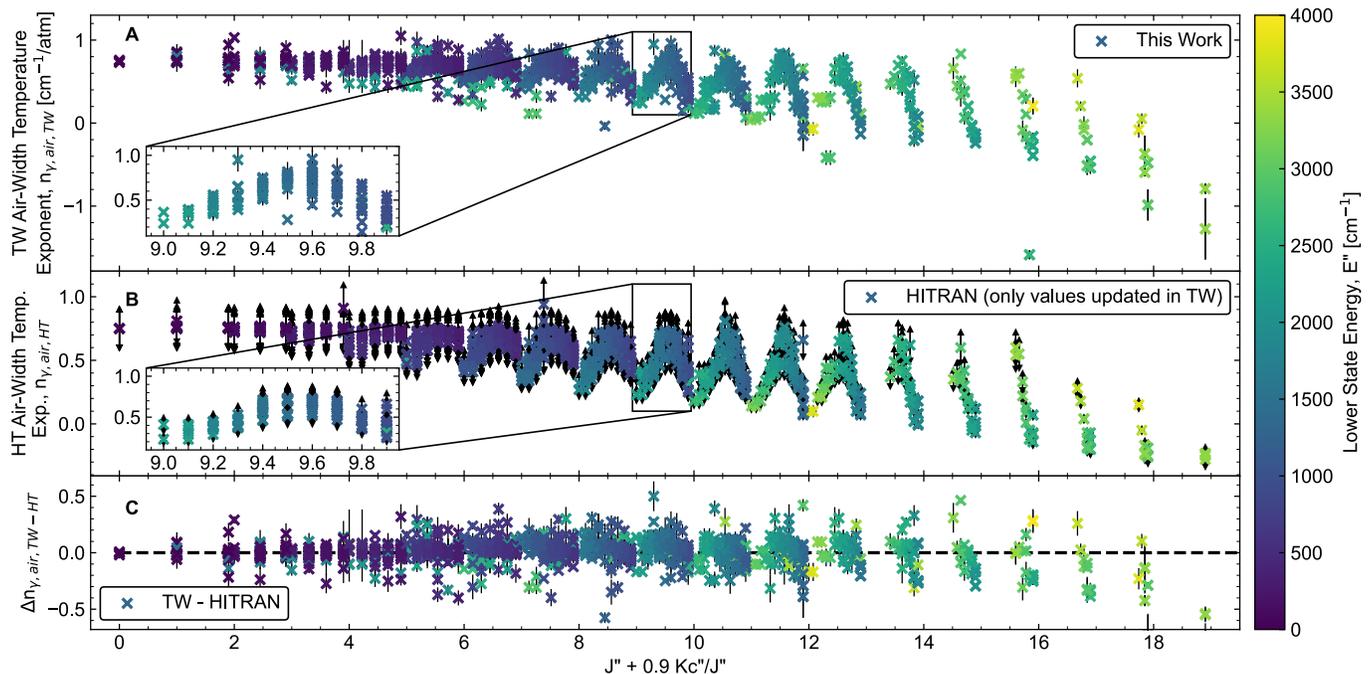

*Figure 8: Measured (panel A) and HITRAN (panel B) TD of the air-broadened width values, $n_{\gamma,air}$, as a function of rotational quantum number J" with a small Kc"/J" offset. Insets in panels A and B highlight the variation of $n_{\gamma,air}$ as a function of Kc" at J" = 9. Panel C shows the difference in $n_{\gamma,air}$ between this work and HITRAN.*

We observe general agreement between $n_{\gamma,air}$ from this work and HITRAN, as shown in Figure 6. For values where $n_{\gamma,air,HT}$ was larger than zero, $n_{\gamma,air,TW} = -0.0255 + 1.06 n_{\gamma,air,HT}$ with a R$^2$ correlation of 0.79. For all 1331 values, $n_{\gamma,air,TW}$ was, on average, 9% larger than $n_{\gamma,air,HT}$. Noting that $n_{\gamma,air,HT}$ does not include SD, we observed that $n_{\gamma,air,HT}$ is, on average, only 2% larger than $n_{\gamma,air,TW,VP}$ (not shown).

Due to the correlation between $n_{\gamma,air}$ and $\gamma^0_{air}$ parameters, it is not possible to know $n_{\gamma,air}$ better than the foundational parameter $\gamma^0_{air}$, as discussed by [23], [35], [50], [51]. As such, the total uncertainty for $n_{\gamma,air}$ was estimated by adding the relative $n_{\gamma,air}$ statistical uncertainty in quadrature with the relative $\gamma^0_{air}$ uncertainty multiplied by $\sqrt{2}\ln(T_{max}/T_{min})^{-1}$, per [23], [50], [51] and as implemented in Part I [13].

Finally, in Figure 9 we compare measured values for $n_{\gamma,air}$ and $\gamma^0_{air}$ from this work. We observe a linear relationship, as shown on the plot, down to approximately $\gamma^0_{air}$ of 0.02. Comparable to Part I for H$_2$O self-widths, $n_{\gamma,air}$ rapidly decreases near $\gamma^0_{air}$ of zero for high-temperature transitions. These transitions are frequently not visible at lower temperatures, such as 300 K, making the ambient temperature reference $\gamma^0_{air}$ an extrapolation that is heavily dependent on the TD parameterization, in this case a single power law $n_{\gamma,air}$. At elevated temperatures, the Doppler width also becomes the dominant contributor to the total width, further obscuring retrievals of $\gamma^0_{air}$ and $n_{\gamma,air}$. Similar to Figure 8, four transitions where $n_{\gamma,air} < -0.6$ are not shown in Figure 9 to best highlight global trends. These transitions can be found in the supplemental database provided with this work.

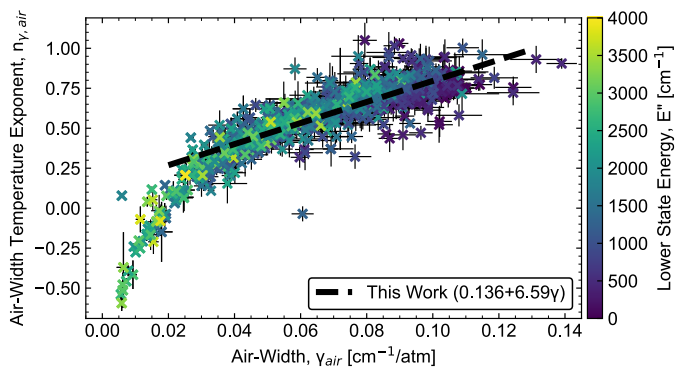

Figure 9: TD of the air-width, $n_{\gamma,air}$, as a function of the air-width, $\gamma^0_{air}$, for all transitions updated in this work. We calculated a trend line between the parameters (for $\gamma^0_{air} > 0.02$) and found a correlation of $R^2 = 0.74$ with the fit parameters shown in the legend.

**Speed-Dependence (SD) of the Air-Width**

SD of the air-width, $a_w$, for the 1024 transitions where this parameter could be retrieved, is shown in Figure 10 as a function of J". Unlike self-width $a_w$ values, which were generally constant, $a_w$ with air as the collisional partner exhibit a positive correlation with J", as shown by the line of best fit with a slope of 0.011J" ($R^2 = 0.3$). While $a_w$ is plotted with a slight Kc" offset, we were not able to identify any sub-structure correlated with Kc" in the $a_w$ parameters, comparable to the insets shown in Figure 7 and Figure 8 for $\gamma^0_{air}$ and $n_{\gamma,air}$, respectively. We calculated $a_w$ values using Equation 9 with an average value of $\alpha$ to represent the average mass of air. We observed similar average values (0.19±0.08 and 0.18±0.11 when retrieved and calculated, respectively). The calculated and the retrieved values show moderate agreement up to approximately J" of 14 when averaging all $n_{\gamma,air}$ values at a given J". Taken individually, calculated values of $a_w$ have a negative $R^2$ value (worse agreement than the mean) with respect to their measured counterparts. While slightly better at predicting $a_w$ for air-$H_2O$ collisions than pure-$H_2O$, our result taken along with other studies [22], [36], [52], [53], [54] identified by Hartmann et al. [44], further supports the hypothesis of Wilzewski et al. that the temperature range, molecule, perturbing molecule, or transition measured impacts the applicability of Equation 9 [54].

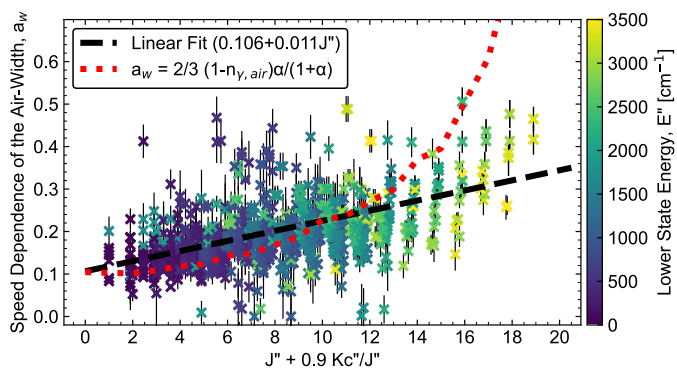

Figure 10: SD of the air-width, $a_w$, as a function of J" with an offset of Kc". A linear best fit to the data is shown in black ($R^2 = 0.3$) to highlight the upward trend in the data. Additionally, measured values of $n_{\gamma,air}$ and Equation 9 were used to calculate $a_w$, with the red line showing the average results for each J" [43].

Like Part I [13], TD of the SD was not included as a unique retrieved parameter in this study. Instead we assume that $a_w$, the ratio of SD and the Lorentz width, is constant with temperature. We calculated $a_w$ at each temperature for 221 strong and isolated features, and observed variations of ±5.2%, which is slightly larger than the 3.9% observed in Part I [13]. Total uncertainties for $a_w$ were calculated with the same technique used for $\gamma^0_{air}$ with an additional ±5.2% included for the lack of TD. For reference, the average statistical uncertainty is ±0.018.

**Air Pressure Shift**

In Figure 11 we show the 858 air pressure shift, $\delta^0_{air}$, values we retrieved in this work using the same adjusted J" x-axis as shown previously. The strong trends with respect to Kc" observed for $n_{\gamma,air}$ and $\gamma^0_{air}$ are not as visible for $\delta^0_{air}$. Panels A, B, and C show $\delta^0_{air}$ measured in this work, $\delta^0_{air}$ in the HITRAN database, and the difference between $\delta^0_{air,TW}$ and $\delta^0_{air,HT}$, respectively.

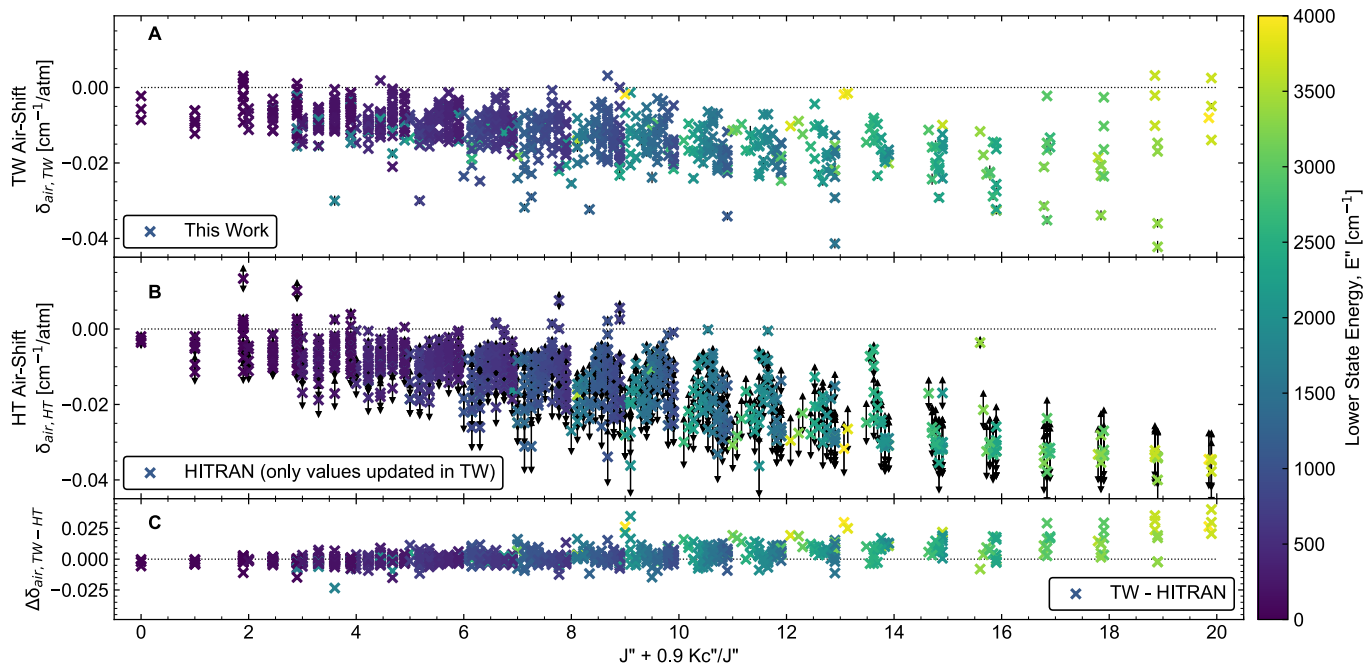

Figure 11: Measured (panel A) and HITRAN (panel B) air-shift, $\delta^0_{air}$, as a function of rotational quantum number J" with a small Kc"/J" offset. Panel C shows the difference in $\delta^0_{air}$ between this work and HITRAN.

From Figure 11, we note the bias between $\delta^0_{air,TW}$ and $\delta^0_{air,HT}$ beginning near J" = 12 and clearly visible in panel C, where $\delta^0_{air,HT}$ is consistently more negative than $\delta^0_{air,TW}$. The majority of HITRAN $\delta^0_{air}$ values updated in this work originate from potential energy surface calculations for $H_2O$-$N_2$ collisions performed by Vispoel et al. in [55] (63% of features) and [56] (22%) that were combined with oxygen data from Gamache and Vispoel for use in the HITRAN database [14]. For $\delta^0_{air}$ taken from these sources, RMS errors relative to this work were 0.008 and 0.005 cm$^{-1}$ for [55] and [56], respectively. In contrast, 14% of the $\delta^0_{air}$ values were taken from a dedicated study of room temperature $H_2O$ and air presented by Toth [48]. The RMS error for these transitions was only 0.001 cm$^{-1}$. Of the 121 Toth transitions updated in this work, only six have double digit J" (maximum J" of 12) because Toth only measured absorption at room temperature. All but one of the HITRAN values for $\delta^0_{air}$ at J" > 12, parameters with the observed bias relative to this work shown in panel C, originate from the potential energy surface calculations of [55] and [56]. The existence of $\delta^0_{air}$ values for these transitions highlights the importance of ab initio calculations for determining parameters that have not been measured, even if those calculations have larger errors than direct measurements. We hope the measurements presented in this work can be used in partnership with future potential energy surface calculations to better predict $\delta^0_{air}$, especially for high-J" $H_2O$ transitions.

Total uncertainty values for $\delta^0_{air}$ were calculated using the same technique as Part I [13], noting that the measured pressure range for $H_2O$ in air is considerably larger than pure $H_2O$ vapor (600 vs 16 Torr). When calculating total uncertainty, we used the same terms identified previously for $\gamma_{air}$ with an additional wavenumber uncertainty term estimated as the DCS wavenumber uncertainty divided by $P_{max}\delta^0_{air}$, the maximum observed shift under the conditions measured (600 Torr or 0.79 atm).

## Temperature-Dependence (TD) of the Air-Shift

Figure 12 shows the 509 retrieved single power law TD of the air-shift values, $n_{\delta,air}$, plotted as a function of J" with a Kc" offset and coloration based on lower state energy comparable to previous figures. In contrast to $n_{\gamma,air}$, we observe $n_{\delta,air}$ increasing as a function of J". Based on the single power law parameterization, a larger $n_{\delta,air}$ at high J" would more agressively attenuate the larger magnitude of $\delta^0_{air}$ for these high-temperature transitions, reducing the observed shift.

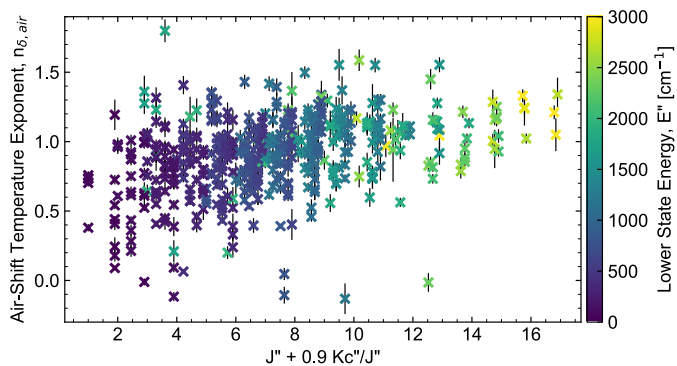

Figure 12: Measured TD of the air-shift, $n_{\delta,air}$, as a function of J" with a 0.9 Kc"/J" offset, colored based on the vibrational assignment of each transition.

In Figure 13 we plot $n_{\delta,air}$ and $\delta^0_{air}$ for the 509 transitions where both were measured, comparable to the inter-comparison of $n_{\gamma,air}$ and $\gamma^0_{air}$ shown previously in Figure 9. Like Figure 9, both $n_{\gamma,air}$ and $n_{\delta,air}$ trend downward to zero or below when $\gamma_{air}$ and $\delta^0_{air}$ approach zero, highlighting the correlation between TD and the reference value.

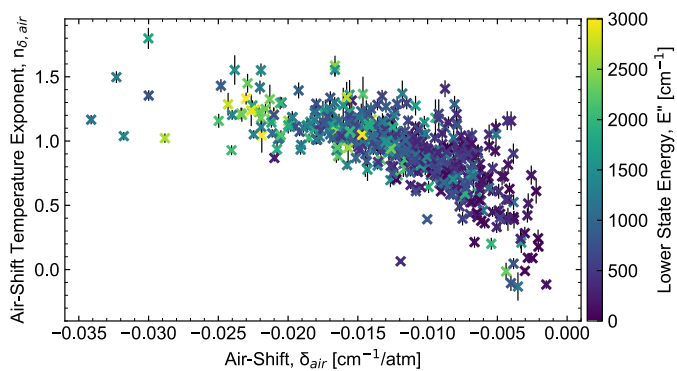

Figure 13: TD of the air-shift, $n_{\delta,air}$, as a function of the air-shift, $\delta^0_{air}$, for all transitions updated in this work.

Total uncertainty was calculated by combining the statistical uncertainty for $n_{\delta,air}$ and that of $\delta^0_{air}$ using the same method of relating uncertainty in $\gamma^0_{air}$ and $n_{\gamma,air}$ shared previously. Using the standalone $\delta^0_{air}$ without incorporating $n_{\delta,air}$ should be done with caution given the interrelated nature of $\delta^0_{air}$ and $n_{\delta,air}$.

## Conclusions

We present 7088 air-broadening parameters for 3366 unique absorption transitions of $H_2O$ between 6600-7650 cm$^{-1}$. These parameters include significant updates to the temperature dependence of the air-shift that are difficult to measure at ambient temperatures and thus have not been broadly catalogued using traditional spectroscopic techniques.

We determine the new line shape parameters using 29 DCS $H_2O$ absorption spectra between 300 and 1300 K and 20 to 600 Torr air, with an average residual noise of 8.0E-4 (absorbance units) and ±1.7E-4 cm$^{-1}$ wavenumber uncertainty. With these measurements, updates are made to transitions with line strengths as low as 3E-31 cm$^{-1}$ / (molecule cm$^{-2}$).

Building on the database presented in Part I, we experimentally determine 3366 $\gamma_{air}^0$, 1331 $n_{\gamma,air}$, and 1000 $a_w$ parameters. We observe good agreement between $\gamma_{air}^0$ and $n_{\gamma,air}$ values measured in this work compared to those in the HITRAN database, which predominantly originate from semi-empirical calculations [14], [45]. We hope these new values for $\gamma_{air}^0$ and $n_{\gamma,air}$ can be used to improve these semi-empirical calculations, especially for high-temperature transitions.

We present 858 updated values for $\delta_{air}^0$ and 509 newly added values for $n_{\delta,air}$. Updated $\delta_{air}^0$ values were shown to agree well with both semi-empirical and experimentally determined HITRAN values up to J" of 14, beyond which a bias is observed in HITRAN values. The inclusion of $n_{\delta,air}$ visibly improved spectral residuals of high-temperature measurements.

In addition to the SDVP values presented, values using the Voigt profile were also retrieved for situations where the fidelity of SD narrowing is not required (e.g. low SNR measurements). Overall, compared to the HITRAN2020 database, spectral residuals across all measured conditions were reduced by a factor of 4.2 in this work using SDVP parameters, improving our understanding of $H_2O$ absorption in air, especially at combustion relevant conditions.

## Declaration of Competing Interests

The authors declare that they have no known competing financial interests or personal relationships that could have appeared to influence the work reported in this paper.

## Acknowledgements

This work was funded through NASA's Exoplanetary Research Program (80NSSC18K0351), the Air Force Research Laboratory (AFRL) (FA8650-20-2-2418), and Air Force Office of Scientific Research (AFOSR) (FA9550-20-1-0328). Portions of this research were performed at the Jet Propulsion Laboratory, California Institute of Technology, under contract with the National Aeronautics and Space Administration and California Institute of Technology. Government sponsorship acknowledged.